\documentclass[12pt,preprint]{revtex4-1}
\usepackage{amsfonts}
\usepackage[final,hiresbb]{graphicx}
\usepackage{amsmath}
\usepackage{amssymb}
\usepackage{amstext}
\usepackage{epstopdf}
\usepackage{makeidx}

\usepackage{graphicx,color}

\makeindex

\begin{document}
\title{Parallelized Hybrid Monte Carlo Simulation of Stress-Induced Texture Evolution}
\author{Liangzhe Zhang$^{a}$, Timothy Bartel$^{b}$, Mark T. Lusk$^{a,}$\footnote{Corresponding author. Tel.: 303-273-3675; fax: 303-273-3919. E-mail: mlusk@mines.edu (M.T. Lusk)}}
\affiliation{$^{a}$Department of Physics, Colorado School of Mines, Golden, CO
80401\\
$^{b}$Sandia National Laboratories, Albuquerque, NM 87185}

\begin{abstract}
A parallelized hybrid Monte Carlo (HMC) methodology is devised to quantify the microstructural evolution of polycrystalline material under elastic loading.  The approach combines a time explicit material point method (MPM) for the mechanical stresses with a calibrated Monte Carlo (cMC) model for grain boundary kinetics. The computed elastic stress generates an additional driving force for grain boundary migration. The paradigm is developed, tested, and subsequently used to quantify the effect of elastic stress on the evolution of texture in nickel polycrystals. As expected, elastic loading favors grains which appear softer with respect to the loading direction. The rate of texture evolution is also quantified, and an internal variable rate equation is constructed which predicts the time evolution of the distribution of orientations.
   
{\bf $Keywords:$} Monte Carlo, grain boundary, elasticity, anisotropy, texture, driving force

{\bf $PACS:$} 81.10.Aj, 64.60.De, 02.60.Cb, 64.60.De, 05.10.Ln
\end{abstract}
\maketitle

\section{Introduction}

The properties of polycrystalline materials, ranging from chemical resistance to fracture toughness, depend on the orientations and populations of the grains of which they are comprised~\cite{Humphreys.2004,Kocks.1998}. When these orientations are non-random, the material is said to have \textit{material texture}~\cite{Dillamore.1964}. This anisotropic feature of materials often develops as a result of manufacturing and associated heat treatment~\cite{Humphreys.2004}. Texture can also result from cyclic thermomechanical loading which causes some orientations to grow at the expense of others. Such cyclic loading is common, for instance, in most electrical devices where it can have a deleterious effect on the solder connections of circuit boards~\cite{Telang.2007,Hofer.1987}. Thermally induced loading leads to elastic stresses which can be partially relieved by the growth of preferentially aligned grains. This type of microstructural evolution, at relatively high temperatures and within the realm of very small deformations, is referred to as \textit{stress-induced texture}.  This texture has not been studied to the same degree as the texture development associated with materials processing~\cite{Dillamore.1964,Hirsch19882863,Lucke1976103,Sarma1998105}, and it is the focus of the work presented here.  Localized yield may also occur, of course, but such plastic deformation is not considered in the current study.

Meso-scale computational interrogations of texture evolution are now commonplace~\cite{Chen.2002,Gurtin.1999,Holm.2001}. We are particularly interested, though, in grain boundary motion in which elastic driving forces are not negligible. Even within this narrow focus,  phase-field~\cite{Chen.2002} and sharp-interface (SI)~\cite{abeyaratne.1990}  approaches have been explored. Missing from the arsenal, though, are Monte Carlo (MC) methods, and this is most likely because a discrete, probabilistic setting does not lend itself to the calculation of stress. The current work remedies this by developing a hybrid algorithm in which a deterministic, continuum computation of elastic energy is dovetailed with MC kinetics for microstructural evolution. In order to focus on the methodology, we restrict our attention to evolution dominated by grain boundary kinetics and disregard independent triple junction kinetics\cite{Gottstein20051535,Gottstein20061065}. This implies that the grain boundary properties are isotropic.  We further assume that the boundary mobility is not a function of its accretive velocity, and so the von Neumann conditions are met~\cite{Gottstein20051535}.  Simulations carried out with a high MC temperature, which is different from the temperature of the physical material, most naturally satisfy these requirements~\cite{Liu.2002, Holm.2001}.

The physical domain is idealized as a continuum with sharp, coherent interfaces with material properties quantized on a cubic grid. MC algorithms generate probabilities for grid point changes in crystal orientation based upon the associated change in global free energy. A quasi-static elastic contribution to the free energy change can therefore be generated by comparing stress states before and after each trial event. We present a numerically efficient means of approximating elastic contribution by merely calculating the stress in a small zone around the trial position. The boundaries of the zone are given constant traction boundary conditions and the work done by the zone on the rest of the domain is accounted for in the free energy shift.   This is implemented for a Potts model~\cite{Wu.1982}. 

A previously developed calibration procedure, here extended from two to three dimensions, is used to endow the hybrid Monte Carlo (HMC) paradigm with physical length, time, and energy scales based on experimentally measurable SI properties~\cite{Liu.2002}.  In order to efficiently implement the approach within a parallelized environment, the computational domain is decomposed into a checkerboard and orientation is updated using a Red/Black (RB) algorithm~\cite{Fried.1990}. An alternative would be to adopt the \emph{N-Fold Way}~\cite{Bortz.1975,Hassold.1993} updating strategy which would dramatically speed up a serial MC calculation but has yet to be parallelized~\cite{Korniss.1999}.

The HMC method is applied to investigate the microstructural evolution of nickel polycrystals under elastic loading. Although our primary task is to develop the meso-scale paradigm, we also show that the data generated can be used to construct a macroscopic kinetic equation for the evolution of the variance in a Gaussian distribution of grain orientations.  This equation can be used to efficiently estimate the effects of elastic loading on material texture.



\section{Grain boundary kinetics within the elastic regime}

SI theory is briefly reviewed because it is used to endow the MC paradigm with physical time and length scales. We focus on a polycrystalline cube of linear elastic, anisotropic material. Away from grain boundaries, the elastic distortion energy, $U$, stress $\mathbf{S}$, and strain $\mathbf{\varepsilon}$ are related:
\begin{equation}
U= \frac{1}{2}\mathbf{S} \cdot \mathbf{\varepsilon}, \quad
\mathbf{S} =\mathbb{C}\mathbf{\varepsilon}.
\label{ela_hooke}
\end{equation}
The fourth order elastic stiffness tensor, $\mathbb{C}$, depends on grain orientation. Under our idealized assumptions, the thermodynamic driving traction for interfacial accretion is~\cite{abeyaratne.1990}:
\begin{equation}
f_{s}=[[U]]-\lessdot \mathbf{S\gtrdot }\cdot \lbrack \lbrack \mathbf{\varepsilon}]]+\kappa \gamma^{*} \ .
\label{driv_trac}
\end{equation}
Here $\gamma^{*} $ is the capillary driving force, $\kappa $ is the local mean curvature.  The expressions $\lessdot \mathbf{\symbol{126}\gtrdot }$ and $[[\symbol{126}]]$ represent domain average and difference across the grain boundary, respectively. The Herring relation~\cite{Herring.1949} can then be used to describe the accretive normal speed of the interface, $v$:
\begin{equation}
v=m_{s}f_{s}\ .  \label{si_kins_1}
\end{equation}
The proportionality constant, $m_{s}$, is grain boundary mobility, which can be measured experimentally along with driving force~\cite{Humphreys.2004}.  The information is then translated into a discrete, probabilistic setting for the prediction of grain boundary motion. Specifically, Equation (\ref{si_kins_1}) is replaced by a fluctuation based flip probability which is carried out at all discrete points on grain borders.

\section{Hybrid Monte Carlo Methodology}
In the absence of elastic stresses, texture development is assumed to be captured by a Q-state Potts model~\cite{Wu.1982} introduced here within a non-dimensional context. The unit cube domain is divided into a uniform $(K\times K\times K)$ lattice. The non-dimensional reference lattice spacing is $\Delta =1/K$. Grain orientation is described with an integer-valued spin field, $q_{i}$, which ranges from $1$ to $Q$. One MC time step is defined as the completion of $N$ trial events $(N=K^{3})$. There is no time scale inherent in the MC model, so it is assigned a time interval of $\tau_{mc}$ for use in subsequent calibration. The system Hamiltonian is assumed to be
\begin{equation}
\mathcal{H}=\sum_{i=1}^{N}\sum_{n=1}^{M}J_{q_{i},q_{n}}\delta _{q_{i},q_{n}}+\sum_{i=1}^{N}b_{i},
\label{H_1}
\end{equation}
where $b_{i}$ is the spatially varying elastic energy associated with the $i^{th}$ lattice, $J_{q_{i},q_{n}}$ is the interaction energy between the two neighbor bins, $M$ is the number of neighbors considered of the selected lattice, and $\delta$ is the Kronecker delta.

Kinetics, within the MC paradigm, is treated via a series of trial changes in the orientation of randomly selected sites. The probability with which such trial \textit{flips} are accepted is related to the overall change in the Hamiltonian of the system. It will be convenient to use the following notation for the change in a field before and after a flip in the phase (spin variable) of one site: $[[\mathcal{H}]]^{\ast }=\mathcal{H}^{Trial}-\mathcal{H}$. Here $\mathcal{H}^{Trial}$ is the Hamiltonian after the flip event. A standard Metropolis algorithm~\cite{Landau.2005} is adopted wherein the probability, $P$, for each flip is a function of the resulting change in Hamiltonian:
\begin{equation}
P=\left\{ \begin{array}{l}e^{-\frac{[[\mathcal{H}]]^{\ast }}{T_{mc}}\text{ }},\text{\ \ \quad }[[\mathcal{H}]]^{\ast } \ \ >0\text{ } \\1,\text{\ \  \ \ \ \ \ \ \ \ \quad }[[\mathcal{H}]]^{\ast }\ \text{ }\leq 0%
\end{array}%
\right. \ .  \label{metrop_1}
\end{equation}
The fundamental MC temperature, $T_{mc}$, parameterizes variability and is not the physical temperature of the material. An issue is how to account for the elastic contribution to $[[\mathcal{H}]]^{\ast }$. 

\begin{figure}[ptb]\begin{center}
\includegraphics[width=0.30\textwidth]{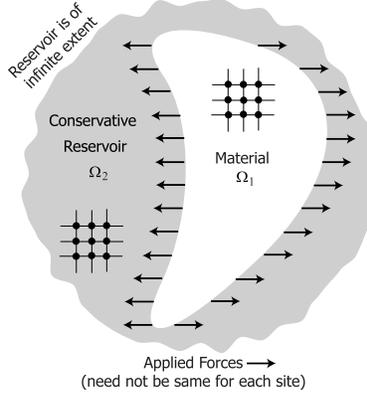}
\caption{
The material domain and its conservative reservoir.
}
\label{conservative_boundary_1}
\end{center}\end{figure}

Suppose that a flip probability is to be calculated at a single discrete point on a grain boundary within a domain of infinite extent. The elastic contribution to $[[\mathcal{H}]]^{\ast }$ is simply the change in the total elastic energy, U, integrated over the entire domain. In principle, this must be calculated by equilibrating the stress twice---once before the flip and once afterwards. This is computationally intractable, and so an approximation is made. First create a neighborhood, $\Omega_1$, around the point and consider the rest of the domain to be a reservoir, $\Omega_2$ (Fig. \ref{conservative_boundary_1}).  A trial flip is assumed to not affect the traction on the boundary of the interior region. Under this assumption, the equilibrated elastic energy which results from a trial flip is the sum of the elastic energy of $\Omega_1$ and the work done by $\Omega_1$ to expand or contract its boundary under constant traction. This boundary motion results from a trial change of elastic stiffness associated with the new orientation being considered. This relation approaches the exact trial energy when $\Omega_1$ increases to include the entire domain. The elastic contribution to $[[\mathcal{H}]]^{\ast }$ is therefore approximated as
\begin{equation}
[[\mathcal{H}]]^{\ast }_{e} = \sum_{\Omega _{1}}U _{i}+\sum_{\partial \Omega _{1}}\vec{T}_{i}\cdot\lbrack \lbrack \vec{u}_{i}]]^{\ast },
\label{H_1}
\end{equation}
where $\vec{T}_{i}$ and $\vec{u}_{i}$ represent traction and displacement associated with the trial flip, respectively.

The most extreme case is to reduce the sub-domain $\Omega _{1}$ to a single point so that
\begin{equation}
\sum_{\partial \Omega _{1}}\vec{T}_{i}\cdot\lbrack \lbrack \vec{u}_{i}]]^{\ast }=-\left\langle \mathbf{\varepsilon }_{p}^{T},\mathbb{C}_{p}
\mathbf{\varepsilon }_{p}\right\rangle +\left\langle \mathbf{\varepsilon }%
_{i}^{T},\mathbb{C}_{i}\mathbf{\varepsilon }_{i}\right\rangle,
\end{equation}
where $p$ and $i$ represent the selected particle and its trial state, respectively. Then we have
\begin{equation}
\lbrack \lbrack \mathcal{H}]]^{\ast }_{e}=-\frac{1}{2}\left\langle \mathbf{\varepsilon }_{p}^{T},\mathbb{C%
}_{p}\mathbf{\varepsilon }_{p}\right\rangle +\frac{1}{2}\left\langle \mathbf{%
\varepsilon }_{i}^{T},\mathbb{C}_{i}\mathbf{\varepsilon }_{i}\right\rangle.
\label{eladriv}
\end{equation}

\section{Time, Length and Energy Scales}
In order to consider the elastic driving force in MC simulations, a correspondence must be established between the parameters of the MC and SI paradigms. This correspondence has been developed for a generalized bulk energy in 2D~\cite{Liu.2002}, and here we extend the approach to a 3D, elastic setting. The characteristic energy, $E_{0}$, length, $l_{0}$, and time, $t_{0}$ are needed to derive expressions for experimentally measurable grain boundary stiffness, $\overline{\gamma^{\ast}}_{s}$,  bulk energy density, $\overline{b}_{s}$, and grain boundary mobility, $\overline{m}_{s}$, in terms of their Monte Carlo counterparts, $\gamma^{\ast}$, $b_{m}$ and $m$:
\begin{align}
\overline{\gamma^{\ast}}_{s}&=\frac{\gamma^{\ast} E_{0}}{ \Delta^{2}J l_{0}^{2}} \notag \\
\overline{b}_{s}& =\frac{2b_{m}E_{0}}{\Delta ^{3}l_{0}^{3}}  \notag \\
\overline{m}_{s}& =\frac{\Delta ^{4}m(\alpha, \varphi)l_{0}^{4}}{J\tau _{mc}E_{0} t_{0}}.
\end{align}
The MC time unit, $\tau _{mc}$, and interaction energy, $J$, are typically set to unity. It is then assumed that both experimental and MC parameters are specified, and the expressions are used to solve the characteristic energy, length, and time scales which link the SI paradigm to a physically meaningful SI setting. A key field in these links is the MC grain boundary mobility, $m(\alpha, \varphi)$, which is calculated using
\begin{equation}
m(\alpha, \varphi)=\frac{m^{\ast}(\alpha)}{\gamma^{\ast}(\alpha,\varphi)}.
\end{equation}
The parameter, $m^{\ast}(\alpha)$, is the MC grain boundary reduced mobility, and it can be measured by studying the capillarity driven shrinkage of internal grains~\cite{Liu.2002}. Since $T_{mc}$ has the units of energy, the nondimensional parameter, $\alpha = \frac{T_{mc}}{J}$, characterizes variability within the MC system. The physical domain size is taken to be the characteristic length, $l_{0}$. The characteristic time, $t_{0}$ is then, 
\begin{equation}
t_{0}=\frac{l_{0}^{2} m^{\ast}(\alpha) \Delta^{2}}{\overline{m}_{s} \overline{\gamma^{\ast}}_{s}},
\label{chatime}
\end{equation}
and the characteristic energy, $E_{0}$, is written as
\begin{equation}
E_{0}=\frac{\overline{\gamma^{\ast}}_{s} l_{0}^{2} \Delta^{2}}{\gamma^{\ast}}.
\label{chaeng}
\end{equation}
Here $\overline{\gamma^{\ast}}_{s}$ is the experimentally measured grain boundary stiffness, and the MC grain boundary stiffness, $\gamma^{\ast}$, can be numerically derived~\cite{Liu.2002}. 

The SI elastic driving force, $\lbrack \lbrack \mathcal{H}]]^{\ast }_{e}$, is then converted into a MC setting using
\begin{equation}
\lbrack \lbrack \mathcal{H}]]^{\ast }_{m}=\frac{\lbrack \lbrack \mathcal{H}]]^{\ast }_{e} \Delta^{3} l_{0}^{3}}{2 E_{0}}.
\label{elamc}
\end{equation}
Here $\lbrack \lbrack \mathcal{H}]]^{\ast }_{m}$ is the elastic driving force in MC format. This expression for elastic driving force is then used in Eq. (\ref{metrop_1}) to give a probabilistic rate of grain boundary motion strictly in terms of SI parameters. This endows the MC paradigm with length, time and energy scales.


\section{Elastic Deformation}

A time-explicit version of Material Point Method (MPM) is used to track elastic deformation~\cite{Sulsky.1996,York.1997}, although other algorithms could be complementary~\cite{Needleman.1985}. Material is viewed as a set of discrete mass points, and a back ground mesh is set to solve the momentum equation. In response to a prescribed loading condition, the mechanical simulator breaks boundary loading into a series of sub-loadings. For each sub-loading, the time-explicit algorithm is launched to seek the quasi-static state in the following manner. Nodal velocity is computed and mapped back to material points. The strain increment associated with each material point is obtained by calculating the gradient of its velocity. Hooke's law is then applied to update stresses of each material point. The state variables of material points are subsequently re-mapped to nodes to start the next iteration. This procedure continues until either the relative stress increment or strain increment criterion is met. The relative stress increment, $\Delta \overline{\sigma },$ and the strain increment, $\Delta \varepsilon,$ of each particle are defined as
\begin{equation}
\Delta \overline{\sigma }_{i}=\frac{\sigma _{i}^{t}-\sigma _{i}^{t-\Delta t}}{\sigma _{i}^{t-\Delta t}},
\end{equation}
and
\begin{equation}
\Delta \varepsilon _{i}=\varepsilon _{i}^{t}-\varepsilon _{i}^{t-\Delta t},
\end{equation}
respectively. Here $i$ represents the particle index, $t$ is the current time step and $(t-\Delta t)$ is the previous time step. The quasi-static state is reached when either the relative stress increment or relative strain increment meets the prescribed convergence criterion for all the particles at time $t$.  

Attention is restricted to cubic symmetry for which there are only three independent elastic constants, $C_{11}$, $C_{12}$ and $C_{44}$.  Grain orientation at each computational grid point determines the principal frame for the elastic stiffness, $\mathbb{C}$:
\begin{equation}
\mathbb{C}=\mathbb{C}_{ijkl}\mathbf{e}_{i}\otimes \mathbf{e}_{j}\otimes \mathbf{e}_{k}\otimes\mathbf{e}_{l}.
\end{equation}
Here $\ \left\{ \mathbf{e}_{i}\right\} $ is an arbitrary Cartesian basis, and the element $\mathbb{C}_{ijkl}$ can be written by the index relationship between matrix and tensor expressions~\cite{Ting.1996}. A rotated stiffness tensor, $\mathbb{C}^{\prime},$ is then given by
\begin{equation}
\mathbb{C}_{ijkl}^{\prime}=\underset{m}{\sum }\underset{n}{\sum }\underset{o}{\sum }\underset{p}{\sum }\mathbb{C}_{ijkl}\mathbf{A}_{im}\mathbf{A}_{jn}\mathbf{A}_{ko}\mathbf{A}_{lp},
\label{stiffrot}
\end{equation}
where $\mathbf{A}$ is a general rotation matrix, which is written as
\begin{equation}
\mathbf{A}=\mathbf{BCD.}
\end{equation}%
The individual rotation matrices, $\mathbf{B,}$ $\mathbf{C,}$\ and $\mathbf{D}$ can be parameterized by three Euler angles, $(\phi ,\theta ,\psi )$~\cite{Arfken.1985}.

\begin{figure}[ptb]\begin{center}
\includegraphics[width=0.30\textwidth]{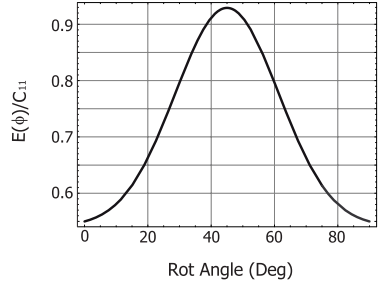}
\caption{The dependence of the Ni Young's modulus on the rotation angle with respect to the z-axis.}
\label{niyoungsmod}
\end{center}\end{figure}

As an example, subsequently considered in more detail, a nickel single crystal is deformed under uniaxial loading with the elastic moduli of $C_{11}=250.8$,  $C_{12}=150.0$ and $C_{44}=123.5$ GPa. The effective Young's modulus along the x-axis of the rotated crystal is depicted in Fig. \ref{niyoungsmod} as a function of crystal rotation about its z-axis.

The static polycrystalline Ni response to elastic loading was then computed. To ensure isotropic behavior, the number of grains, $N_{g}$, was determined using~\cite{Nygards.2003}
\begin{equation}
N_{g}=\frac{0.03}{\epsilon_{\rm iso}^{2}}(\eta-1),  \label{Nygards}
\end{equation}
where $\epsilon_{\rm iso}$ is the acceptable relative error of calculated stress and strain among independent simulations, and $\eta$ is the characteristic parameter of anisotropy:
\begin{equation}
\eta=\frac{2C_{44}}{C_{11}-C_{12}}.
\end{equation}
This implies that isotropy along with a $2\%$ error will be achieved for a nickel polycrystal with 300 grains. The experimental  value of the Young's modulus of polycrystalline Ni is 225 GPa~\cite{Ledbetter.1982}, while our MPM calculation gives 221 GPa. The difference is $1.8\%$. The agreement between simulations and experiments allows us to apply the technique to a polycrystalline setting in which the grain boundaries evolve.

\section{Microstructural Evolution}
In typical implementations of the Q-state Potts simulations, updating follows immediately after a trial event is accepted.  This amounts to a Gauss-Seidel (GS) update~\cite{Wu.1982}. The efficiency of this approach can be remarkably improved using the N-Fold Way algorithm~\cite{Hassold.1993}. When applied within a parallel environment, though, communication between processors will be asynchronous. Since we seek to develop a simulation approach amenable to large-scale, parallelized environments, we give up the N-Fold Way in favor of a Red/Black (RB) updating scheme~\cite{Fried.1990}. In the RB scheme, an MC domain is decomposed into a checkerboard. Since each spin site interacts only with its four nearest neighbors, the outcome of a trial event is determined only by the four opposite colored lattice sites. The update of the selected particle will not affect any other particle with a different color. This allows the particles updating in parallel. One MC step is therefore split into black/red and red/black half steps. In each, the trial states of the red or black particles are not updated until this half step finishes. Only two updating steps are required in RB algorithm as opposed to a series of updates associated with GS algorithm.  Orientations on the boundaries of the decomposed domain require a communication between processors. 

The RB updating approach is validated by recovering the predictions of \emph{Isotropic Grain Growth Theory}~\cite{Humphreys.2004,Holm.2001,Holm.1993,Tikare.1998} which predicts that the mean grain radius at time t is $R=c_{1}t^{1/n}$. Here $c_{1}$ is a constant and $n$ is referred to as the \textit{grain growth exponent}. Within the current setting, $n=2$~\cite{Humphreys.2004}. Therefore, grain volume, $V$, is predicted to evolve as $V^{3/2} =c_{2}t$, where $c_{2}$ is a constant.

\begin{figure}[ptb]\begin{center}
\includegraphics[width=0.30\textwidth]{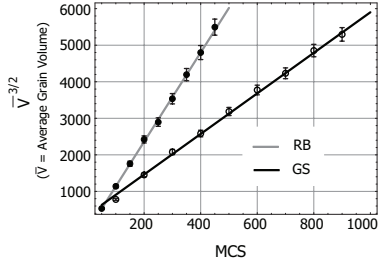}
\caption{Relationships between rate of change of average grain size and MC time for two updating algorithms. The results pertain to a 3D polycrystal. The solid gray line and the solid black lines are fits to the GS and RB data, respectively. Grid size=$60^{3}$ and $\alpha=1.0.$ Error bars are for the standard deviation of the $10$ averaged simulation results. Volume is measured as the number of cubic grid cells comprising a grain.}
\label{3d_growth_01}
\end{center}\end{figure}

A set of three-dimensional simulations were considered on a $60^{3}$ grid, wherein a total of $99$ orientations were randomly distributed. Grain boundaries were all given the same interaction energy, and the microstructure was subsequently allowed to evolve with a variability parameter $\alpha=1.0$. The average grain size was collected over time and is shown in Fig.\ref{3d_growth_01}.  The data indicate that both RB and GS have the correct power law behavior, although the mobility of the RB algorithm is higher than that for GS. This is because the first half step updating in RB algorithm creates more boundary corrugation and the capillarity effect is therefore accelerated.

\section{Results}

The HMC algorithm was first analyzed within an idealized one-dimensional setting. A bi-crystalline bar of length, $L$, and cross-sectional area, $A$, was fixed at left and loaded with constant traction, $F$, at right. The Young's moduli of the two domains at left and right are $E_{1}$ and $E_{2}$ $(E_{1}>E_{2})$, respectively. The domain is discretized by $100$ lattice sites $(N=100)$. Eq. (\ref{eladriv}) was then used to calculate the elastic driving force:
\begin{equation}
\lbrack \lbrack \mathcal{H}]]^{\ast }_{e}=\pm \frac{F^{2}}{2 A^{2}}(\frac{1}{E_{1}}-\frac{1}{E_{2}}) .
\label{1ddri}
\end{equation}
Positive driving force is obtained if a particle of grain two is selected and flipped to grain one, and the trial flip is therefore accepted. This implies the grain boundary moves to the left side with a speed of one particle distance per MC step, i.e.,
\begin{equation}
v=\frac{L}{N\tau _{mc}}.
\end{equation}
The HMC implementation delivered a grain boundary velocity (averaged over $100$ simulations) which is consistent with this analytical solution, and the relative error was found to be less than one percent. Note that the grain boundary moves so as to grow the softer grain. This is the guiding rule for texture evolution in response to elastic loading; softer orientations will grow at the expense of their stiffer counterparts. Here \emph{stiffer} means that the grain offers a higher effective Young's modulus along the axis of loading.  Of course, both elastic stored energy and work are considered. This mechanism is different from the approaches which consider the elastic stored energy only~\cite{Rad.2009}.

A series of three-dimensional simulations were subsequently carried out. One $mm^{3}$ polycrystalline nickel cube was deformed at $450\,^{\circ}\mathrm{C}$. For computational convenience, in later calibration, grain boundary energy and grain boundary mobility are taken to be isotropic, although in reality the two items are characterized by anisotropy~\cite{Humphreys.2004,Olmsted20093694}. Room temperature elastic moduli were adopted even though there is a~$8 \%$ difference between room temperature and $450\,^{\circ}\mathrm{C}$ values~\cite{Ledbetter.1982}. One million computational particles were used with each assigned a random initial orientation. Only the rotation angle with respect to the z-direction was varied to more clearly interpret the results. The other two Euler angles were set to zero. The rotation angle was varied from $0^{\circ }$ to $90^{\circ }$ in $1^{\circ }$ steps; therefore $91$ orientations were used in total. Fig. \ref{niyoungsmod} shows the orientation effect on the Young's modulus in the x-direction. The $45^{\circ }$ is the stiffest orientation, while $0^{\circ}$ and $90^{\circ }$\ are the softest orientations. The random initial microstructure was first allowed to coarsen over $100$ MC steps at $\alpha=2.6$ to construct a polycrystalline microstructure. A uniaxial tension of $30$ MPa was applied in the x-direction in $75$ loading substeps. In each sub-step, the quasi-static elastic state was obtained and was then followed by one MC step during which the microstructure evolved. In general, texture evolution is described by two mechanisms, grain boundary kinetics or triple junction kinetics~\cite{Gottstein20061065}. In order to more cleanly isolate the influence of elastic loading, we focus only on the former and thus implicitly assume that grain boundary properties are isotropic. As previously noted, we also assume that the the mobility is not a function of accretive velocity~\cite{Gottstein20051535}. Within the MC setting, these conditions are met by restricting attention to high MC temperatures. The constant, experimentally reasonable, grain boundary energy~\cite{Humphreys.2004,Olmsted20093694} and grain boundary mobility~\cite{Saindrenan.1999} are assigned, although the energy and mobility are known to depend on the details of grain boundary structure~\cite{Humphreys.2004}. Key boundary properties and their corresponding MC parameters, obtained from Eq. (\ref{chatime}) and Eq. (\ref{chaeng}),  are listed in Table \ref{chapara}. 
\begin{table}[table1] 
\centering
\begin{tabular}{|c|c|c|}
\hline
Parameters & Values \\ \hline
$m^{\ast }$  & 0.27  \\ \hline
$\gamma^{\ast}$ & 2.4  \\ \hline
$\overline{m}_{s}$ $(m^{4}/J.s)$  & $1.0\times 10^{-13}$ \\ \hline
$\overline{\gamma^{\ast }}_{s}$ $(J/m^{2})$ & 0.45  \\ \hline
$\overline{m^{\ast }}$ $(m^{2}/s)$ & $0.45\times 10^{-13}$  \\ \hline
$l_{0}$ $(m)$ & $1.0 \times10^{-3}$  \\ \hline
$t_{0}$ $(s)$ & 600  \\ \hline
$E_{0}$ $(J)$ &  $1.875 \times10^{-11}$ \\ \hline
\end{tabular}
\caption{Grain boundary properties and required  characteristic parameters to link MC and SI models.}
\label{chapara}
\end{table}

\begin{figure}[ptb]\begin{center}
\includegraphics[width=0.30\textwidth]{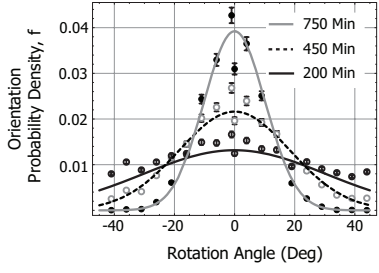}
\caption{Evolution of texture under elastic loading with a single Euler angle description of each orientation. Every five of $91$ simulation data are shown as discrete points along with Gaussian fits as solid curves. Results are averaged over $10$ independent runs. Errors bars quantify the standard deviation.}
\label{texture_ela_01}
\end{center}\end{figure}

Since any sufficiently high MC temperature is valid, we chose a value such that one MC step is equal to $10$ minutes. Fig. \ref{texture_ela_01} shows the Young's modulus probability density associated with individual orientations at three time slices. Consistent with intuition and the results of our 1D study, evolution favors grains which are softer with respect to the elastic loading axis. 
To quantify the elastic effect on texture evolution, the texture histograms were fitted to a time dependent Gaussian equation:
\begin{equation}
f(\phi ,t)= \frac{1}{g(t)\sqrt{2\pi}  \rm{\it Erf}(45/(\sqrt{2}g(t)))}Exp[-\frac{\phi ^{2}}{2g(t)^{2}}],
\label{text_evl}
\end{equation}
where $f$ is the orientation probability density, $\phi$ is the orientation angle, and $\it Erf$ is an error function. The evolution of texture is thus distilled to a single variance, g. In order to characterize this process with a single parameter, the rate of change of the variance is assumed to be proportional to the current variance,
\begin{equation}
\frac{dg}{dt}=-\frac{g}{\tau },  \label{varn_eqn}
\end{equation}
where $\tau$ is a characteristic time constant. Since Equation (\ref{1ddri}) indicates that the driving force is proportional to $\sigma^{2}$, the solution to Equation (\ref{varn_eqn}) is written as
\begin{equation}
g(t)=\beta\sigma ^{2} e^{-\frac{t}{\tau}},
\end{equation}
where the time constant, $\tau$, and coefficient, $\beta$, were then fitted to the HMC data: $\tau=500 \text{ }min$ and $\beta= 0.05 \text{ } ^{\circ}/\rm{\it{MPa}}^{2}$. Within our restrictive ansatz, the characteristic time, $\tau$, describes the rate of texture evolution in polycrystalline Ni under a uni-axial loading. This fitted Gaussian variance, $g(t)$, was subsequently used in Eq. (\ref{text_evl}) to produce the solid curves of Fig. \ref{texture_ela_01}.  
\begin{figure}[ptb]\begin{center}
\includegraphics[width=0.30\textwidth]{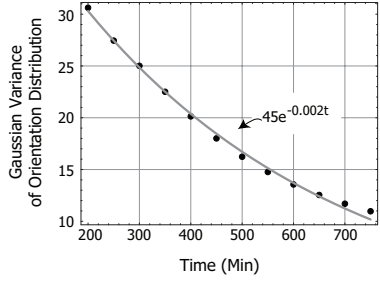}
\caption{Evolution of the Gaussian variance for the distribution of orientations. The solid dots are the HMC results while the gray curve is the solution of Eq. (\ref{varn_eqn}).}
\label{variance}
\end{center}\end{figure}

\begin{figure}[ptb]\begin{center}
\includegraphics[width=0.30\textwidth]{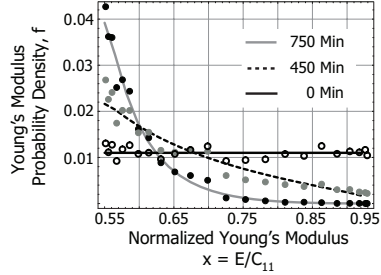}
\caption{Re-plot of texture data of Fig. \ref{texture_ela_01} with horizontal axis now the effective Young's modulus.}
\label{texture_youngs_91}
\end{center}\end{figure}

More general polycrystalline settings can not use a single angle of misorientation to describe texture, so we now show a second approach which employs the effective Young's modulus as a texture measure. First the Young's modulus data of Fig. \ref{niyoungsmod} are fitted to a Gaussian function:
\begin{equation}
x(\phi)=a\text{\\ }Exp[-\frac{(\phi-45^{\circ})^{2}}{2b^{2}}]+c,
\label{young_gau} 
\end{equation}
where $x(\phi)=E(\phi)/C_{11}$, $a =0.388$, $b =16.40^{\circ }$, and $c =0.541$. It is the inverse of this relation, though, that is required in order to describe the texture in terms of the Young's modulus:
\begin{equation}
\phi(x)=45^{\circ}-[-2b^{2}Log(\frac{x-c}{a})]^{\frac{1}{2}},\text{ \ \ } 0.55<x<0.929.
\label{inv_young} 
\end{equation}
The orientation probability density, f, can now be written as a function of Young's modulus by combining Equations (\ref{inv_young}) and (\ref{text_evl}), and the result is plotted in Fig. \ref{texture_youngs_91}. The distribution, now referred to as the Young's modulus probability density, is initially uniform but favors softer grains with increasing time. 

\begin{figure}[ptb]\begin{center}
\includegraphics[width=0.30\textwidth]{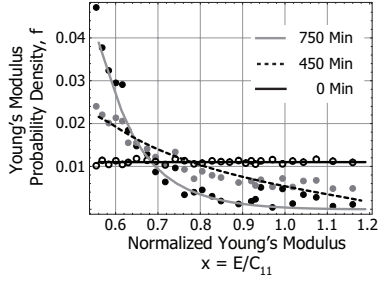}
\caption{Evolution of texture under elastic loading with three arbitrary Euler angles description of each orientation. The $999$ data points are integrated into $91$ points, and every three of the $91$ data points are shown as discrete points along with Gaussian fits as solid curves. Results are averaged over $20$ independent runs. Errors bars quantify the standard deviation.}
\label{texture_youngs_999}
\end{center}\end{figure}

The previous numerical experiment used only one Euler angle in order to make the study particularly transparent. A more general application was therefore considered in which $999$ random orientations span all three Euler angles. Fig. \ref{texture_youngs_999} shows the resulting texture evolution quantified by the effective Young's modulus. The data are fitted to the composition of functions given in Equations (\ref{inv_young}) and (\ref{text_evl}) with $a=0.667$, $b =16.40^{\circ }$, and $c =0.541$. Note that only the parameter $a$ differs from its value for a single angle description of orientation, and the previously obtained Gaussian variance, g, is utilized.
\section{Discussion}
The hybrid Monte Carlo (HMC) paradigm combines a deterministic, continuum elasticity solver with a probabilistic, discrete algorithm for grain boundary evolution. The paradigm is intended for application where elastic loading influences grain boundary motion on time scales sufficiently slow that inertial effects can be neglected. 

The isothermal, uniaxial loading of polycrystalline Ni is adopted in to demonstrate the idea. Attention was restricted to a MC temperature regime for which the grain boundary properties are isotropic and variations in grain orientation were initially restricted to single Euler angle. In line with intuition, it was found that grains which are softer with respect to the loading direction are energetically favored. The results also show that the texture conforms to a Gaussian histogram with a variation that can be described by a first order differential equation. For quantifying the texture evolution where more general Euler angles are allowed, the independent variable can be changed from angle to the effective Young's modulus with respect to loading direction. 

The HMC methodology was then applied to a more realistic Ni polycrystal in which all three Euler angles were used. As expected, the softer grains with respect to the loading direction eventually dominate the polycrystalline media. The data were accurately fitted to the same relation used for a single angle of orientation. The fit describes the rate at which texture evolves using a single characteristic time. Significantly, characteristic time at which the distribution tightens was found to be the same as for the case where only a single angle of orientation was allowed to vary. An appropriate next step is to compare the predictions of this model directly with experimental data for texture evolution under uniaxial loading. We are not aware of any such data with which we can currently make such a comparison. In all implementations, only the most primitive approximation was used for the elastic contribution to energy changes do to re-orientation. The domain over which stress equilibration is considered could be extended to a larger neighborhood of trial flip sites.

This hybrid computational methodology for studying texture evolution offers an alternative to sharp-interface and phase-field modeling. The resulting texture evolution maps are expected to be useful in materials and process design as well as in part performance assessment throughout service life. An extension of this work to include linearized inelastic deformation, currently underway, will allow the influence of linear plastic deformation on texture to be accounted for as well. The approach can also be extended to account for an evolving temperature field via coupling with the heat equation. 

\section{Acknowledgments}

The research is supported by Sandia National Laboratories and is guided by Elizabeth Holm. Sandia National  Laboratories are operated by the Sandia Corporation, a Lockheed Martin Company, for the United States Department of Energy under contract DE-AC04-94AL85000.  We also acknowledge the Golden Energy Computing Organization at the Colorado School of Mines for the use of resources acquired with financial assistance from the National Science Foundation and the National Renewable Energy Laboratories.

\section{Appendix. Numerical Error Analysis}
\begin{figure}[ptb]\begin{center}
\includegraphics[width=0.45\textwidth]{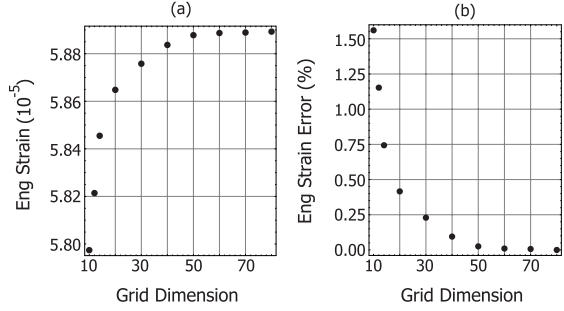}
\caption{Uniaxial tension test results of a 2D polycrystal: (a) computed engineering strains and (b) the relative errors of the engineering strains with respect to grid size.}
\label{hex_engstrn}
\end{center}\end{figure}

The numerical errors associated with MC simulations have been comprehensively studied~\cite{Landau.2005}, so here we focus on errors associated with the mechanical response. Imbalances in force can be made arbitrarily small by tightening the convergence criteria~\cite{Scarborough.1955}. In practice, though, limits should be set which balance the desire to reduce such \textit{numerical errors} with the need to obtain an efficient algorithm. The two interior convergence parameters of relative stress increment and strain increment were set to $1.0\times 10^{-9}$ and $1.0\times10^{-5},$ respectively. A second class of numerical errors is associated with the background mesh size and the number of computational particles per cell. To quantify these, uniaxial tension tests were carried out on a polycrystalline system with seven columnar, hexagonal grains. The total number of computational particles was varied but the number of particles per cell was fixed at four, i.e., two particles per cell in each direction in a given cross-section.  Figs. \ref{hex_engstrn} (a) and (b) demonstrate the calculated engineering strains and relative errors, respectively.  The minimum number of particles required to meet a $1\%$ relative error was determined to be $12\times10$. This analysis provides a criterion to determine the number of particles required to represent one grain in polycrystalline media. A direct extension to 3D implies that $900$ computational particles should be used per grain, and this rule is followed in all simulations.

\section{references}

\begin{thebibliography}{10}

\bibitem{Humphreys.2004}
F.~J. Humphreys and M.~Hatherly.
\newblock {\em Recrystallization and Related Annealing Phenomena}.
\newblock Elsevier Ltd, Oxford, United Kingdom, 2004.

\bibitem{Kocks.1998}
U.~F. Kocks, C.~N. Tome, and H.-R. Wenk.
\newblock {\em Texture and Anisotropy}.
\newblock Cambridge University Press, Cambridge, United Kingdom, 1998.

\bibitem{Dillamore.1964}
I.~L. Dillamore and W.~T. Roberts.
\newblock Rolling textures in f.c.c. and b.c.c. metals.
\newblock {\em Acta Metallurgica}, 12(3):281 -- 293, 1964.

\bibitem{Telang.2007}
A.~U. Telang, T.~R. Bieler, A.~Zamiri, and F.~Pourboghrat.
\newblock Incremental recrystallization/grain growth driven by elastic strain
  energy release in a thermomechanically fatigued lead-free solder joint.
\newblock {\em Acta Materialia}, 55(7):2265 -- 2277, 2007.

\bibitem{Hofer.1987}
G.~Hofer.
\newblock {Texture Dependent Young's Modulus in Austenitic Cladding}.
\newblock {\em Textures and Microstructures}, 8-9:611--617, 1987.

\bibitem{Hirsch19882863}
J.~Hirsch and K.~Lucke.
\newblock Overview no. 76: Mechanism of deformation and development of rolling
  textures in polycrystalline f.c.c. metals--i. description of rolling texture
  development in homogeneous cuzn alloys.
\newblock {\em Acta Metallurgica}, 36(11):2863 -- 2882, 1988.

\bibitem{Lucke1976103}
K.~Lucke, R.~Rixen, and M.~Senna.
\newblock Formation of recrystallization textures in rolled aluminum single
  crystals.
\newblock {\em Acta Metallurgica}, 24(2):103 -- 110, 1976.

\bibitem{Sarma1998105}
G.~B. Sarma, B.~Radhakrishnan, and T.~Zacharia.
\newblock Finite element simulations of cold deformation at the mesoscale.
\newblock {\em Computational Materials Science}, 12(2):105 -- 123, 1998.

\bibitem{Chen.2002}
L.~Q. Chen.
\newblock {Phase-Field Models for Microstructure Evolution}.
\newblock {\em Annual Review of Material Research}, 32:113--140, 2002.

\bibitem{Gurtin.1999}
M.~E. Gurtin and M.~T. Lusk.
\newblock Sharp-interface and phase-field theories of recrystallization in the
  plane.
\newblock {\em Physica D: Nonlinear Phenomena}, 130(1-2):133 -- 154, 1999.

\bibitem{Holm.2001}
E.~A. Holm, G.~N. Hassold, and M.~A. Miodownik.
\newblock On misorientation distribution evolution during anisotropic grain
  growth.
\newblock {\em Acta Materialia}, 49(15):2981 -- 2991, 2001.

\bibitem{abeyaratne.1990}
R.~Abeyaratne and J.~K. Knowles.
\newblock On the driving traction acting on a surface of strain discontinuity
  in a continuum.
\newblock {\em Journal of the Mechanics and Physics of Solids}, 38(3):345 --
  360, 1990.

\bibitem{Gottstein20051535}
G.~Gottstein, Y.~Ma, and L.S. Shvindlerman.
\newblock Triple junction motion and grain microstructure evolution.
\newblock {\em Acta Materialia}, 53(5):1535 -- 1544, 2005.

\bibitem{Gottstein20061065}
G.~Gottstein and L.S. Shvindlerman.
\newblock Grain boundary junction engineering.
\newblock {\em Scripta Materialia}, 54(6):1065 -- 1070, 2006.

\bibitem{Liu.2002}
P.~Liu and M.~T. Lusk.
\newblock {Parametric links among Monte Carlo, phase-field, and sharp-interface
  models of interfacial motion}.
\newblock {\em Phys. Rev. E}, 66:061603, 2002.

\bibitem{Wu.1982}
F.~Y. Wu.
\newblock {The Potts model}.
\newblock {\em Reviews of Modern Physics}, 54(1):235--268, 1982.

\bibitem{Fried.1990}
H.~Fried.
\newblock {The checkerboard update Glauber model, cellular automata and Ising
  models}.
\newblock {\em J. Phys. A}, 23:4165--4181, 1990.

\bibitem{Bortz.1975}
A.~B. Bortz, M.~H. Kalos, and J.~L. Lebowitz.
\newblock A new algorithm for monte carlo simulation of ising spin systems.
\newblock {\em Journal of Computational Physics}, 17(1):10 -- 18, 1975.

\bibitem{Hassold.1993}
G.~N. Hassold and E.~A. Holm.
\newblock {A fast serial algorithm for the finite temperature quenched Potts
  model}.
\newblock {\em Computers in Physics}, 7(1):97--107, 1993.

\bibitem{Korniss.1999}
G.~Korniss, M.~A. Novotny, and P.~A. Rikvold.
\newblock Parallelization of a dynamic monte carlo algorithm: A partially
  rejection-free conservative approach.
\newblock {\em Journal of Computational Physics}, 153(2):488 -- 508, 1999.

\bibitem{Herring.1949}
C.~Herring.
\newblock {\em Surface tension as a motivation for sintering}.
\newblock McGraw-Hill, New York, 1949.

\bibitem{Landau.2005}
D.~Landau and K.~Binder.
\newblock {\em A Guide to Monte Carlo Simulations in Statistical Physics}.
\newblock Cambridge University Press, Cambridge, United Kingdom, 2005.

\bibitem{Sulsky.1996}
D.~Sulsky and H.~L. Schreyer.
\newblock Axisymmetric form of the material point method with applications to
  upsetting and taylor impact problems.
\newblock {\em Computer Methods in Applied Mechanics and Engineering},
  139(1-4):409 -- 429, 1996.

\bibitem{York.1997}
A.~R.~York II.
\newblock {\em Development of Modifications to The Material Point Method for
  the Simulation of Thin Membranes, Compressible Fluids and Their
  Interactions}.
\newblock PhD thesis, Sandia National Laboratories, 1997.

\bibitem{Needleman.1985}
A.~Needleman, R.~J. Asaro, J.~Lemonds, and D.~Peirce.
\newblock {Finite Element Analysis of Crystalline Solids}.
\newblock {\em Computer Methods in Applied Mechanics and Engineering},
  52:689--708, 1985.

\bibitem{Ting.1996}
T.~C.~T. Ting.
\newblock {\em Anisotropic Elasticity: Theory and Applications}.
\newblock Oxford University Press, United States, 1996.

\bibitem{Arfken.1985}
G.~Arfken.
\newblock {\em Mathematical Methods for Physicists}.
\newblock Academic Press, Orlando, FL, 1985.

\bibitem{Nygards.2003}
M.~Nygards.
\newblock {Number of grains necessary to homogenize elastic materials with
  cubic symmetry}.
\newblock {\em Mechanics of Materials}, 35:1049--1057, 2003.

\bibitem{Ledbetter.1982}
H.~M. Ledbetter.
\newblock {Temperature behaviour of Young's moduli of forty engineering
  alloys}.
\newblock {\em Cryogenics}, pages 653--656, 1982.

\bibitem{Holm.1993}
E.~A. Holm, D.~J. Srolovitz, and J.~W. Cahn.
\newblock {Microstructual evolution in two-dimensional two-phase polycrystals}.
\newblock {\em Acta Materialia}, 41(4):1119--1136, 1993.

\bibitem{Tikare.1998}
V.~Tikare, E.~A. Holm, D.~Fan, and L.Q. Chen.
\newblock {Comparison of phase-field and Potts models for coarsening
  processes}.
\newblock {\em Acta Materialia}, 47(1):363--371, 1998.

\bibitem{Rad.2009}
B.~Radhakrishnan and G.~Sarm.
\newblock {Coupled Finite Element-Potts Model Simulations of Grain Growth in
  Copper Interconnects}.
\newblock {\em Mater. Res. Soc. Symp. Proc.}, 1156-1161, 2009.

\bibitem{Olmsted20093694}
D.~L. Olmsted, S.~M. Foiles, and E.~A. Holm.
\newblock Survey of computed grain boundary properties in face-centered cubic
  metals: I. grain boundary energy.
\newblock {\em Acta Materialia}, 57(13):3694 -- 3703, 2009.

\bibitem{Saindrenan.1999}
R.~Le Gall, G.~Liao, and G.~Saindrenan.
\newblock {Experimental Determinaton of Nickel Grain Boundary Mobility During
  Recrystallization}.
\newblock {\em Materials Science Forum}, 294-296:509--512, 1999.

\bibitem{Scarborough.1955}
J.~B. Scarborough.
\newblock {\em Numerical Mathematical Analysis}.
\newblock Oxford University Express, Oxford, London, 1955.

\end{thebibliography}

\end{document}